\documentclass[longauth]{aa}

\usepackage{graphicx}
\usepackage{graphics}

\usepackage{longtable}
\usepackage{txfonts}

\usepackage{epsf,epsfig}
\usepackage[dvips]{color}
\usepackage{color}
\usepackage{amssymb,amsfonts}
\usepackage{latexsym}
\usepackage{float}
\usepackage{tabularx}
\usepackage{multirow}
\usepackage{natbib}

\mathchardef\mhyphen="2D
\bibpunct{(}{)}{;}{a}{}{,} % to follow the A&A style

\begin{document}

\title{Probing the extent of the non-thermal emission from the Vela~X
  region at TeV energies with H.E.S.S.}

\subtitle{}

\author{H.E.S.S. Collaboration
\and A.~Abramowski \inst{1}
\and F.~Acero \inst{2}
\and F.~Aharonian \inst{3,4,5}
\and A.G.~Akhperjanian \inst{6,5}
\and G.~Anton \inst{7}
\and S.~Balenderan \inst{8}
\and A.~Balzer \inst{7}
\and A.~Barnacka \inst{9,10}
\and Y.~Becherini \inst{11,12}
% ...
}

\institute{
Universit\"at Hamburg, Institut f\"ur Experimentalphysik, Luruper Chaussee 149, D 22761 Hamburg, Germany \and
Laboratoire Univers et Particules de Montpellier, Universit\'e Montpellier 2, CNRS/IN2P3,  CC 72, Place Eug\`ene Bataillon, F-34095 Montpellier Cedex 5, France \and
Max-Planck-Institut f\"ur Kernphysik, P.O. Box 103980, D 69029 Heidelberg, Germany \and
Dublin Institute for Advanced Studies, 31 Fitzwilliam Place, Dublin 2, Ireland \and
National Academy of Sciences of the Republic of Armenia, Yerevan  \and
Yerevan Physics Institute, 2 Alikhanian Brothers St., 375036 Yerevan, Armenia \and
Universit\"at Erlangen-N\"urnberg, Physikalisches Institut, Erwin-Rommel-Str. 1, D 91058 Erlangen, Germany \and
University of Durham, Department of Physics, South Road, Durham DH1 3LE, U.K. \and
Nicolaus Copernicus Astronomical Center, ul. Bartycka 18, 00-716 Warsaw, Poland \and
CEA Saclay, DSM/Irfu, F-91191 Gif-Sur-Yvette Cedex, France \and
APC, AstroParticule et Cosmologie, Universit\'{e} Paris Diderot, CNRS/IN2P3, CEA/Irfu, Observatoire de Paris, Sorbonne Paris Cit\'{e}, 10, rue Alice Domon et L\'{e}onie Duquet, 75205 Paris Cedex 13, France,  \and
Laboratoire Leprince-Ringuet, Ecole Polytechnique, CNRS/IN2P3, F-91128 Palaiseau, France
% ...
}

\offprints{\\
florent.dubois@lapp.in2p3.fr,\\
bernhard.glueck@physik.uni-erlangen.de}

\date{\today}

\abstract
{Vela~X is a region of extended radio emission in the western part of the Vela constellation: one of the nearest pulsar wind nebulae (PWNe), and associated with the energetic Vela pulsar (PSR B0833-45). Extended very-high-energy (VHE) $\gamma$-ray emission (HESS $\mathrm{J0835\mhyphen 455}$) was discovered using the H.E.S.S. experiment in 2004. The VHE $\gamma$-ray emission was found to be coincident with a region of X-ray emission discovered with ${\it ROSAT}$ above 1.5 keV (the so-called \textit{Vela~X cocoon}): a filamentary structure extending southwest from the pulsar to the centre of Vela~X.}
{A deeper observation of the entire Vela~X nebula region, also including larger offsets from the cocoon, has been performed with H.E.S.S.
This re-observation was carried out in order to probe the extent of the non-thermal emission from the Vela~X region at TeV energies and to investigate its spectral properties.}
{In order to increase the sensitivity to the faint $\gamma$-ray emission from the very extended Vela~X region, a multivariate analysis method combining three complementary reconstruction techniques of Cherenkov-shower images is applied for the selection of $\gamma$-ray events. The analysis is performed with the On/Off background method, which estimates the background from separate observations pointing away from Vela~X; towards regions free of $\gamma$-ray sources but with comparable observation conditions.}
{The $\gamma$-ray surface brightness over the large Vela~X region reveals that the detection of non-thermal VHE $\gamma$-ray emission from the PWN HESS $\mathrm{J0835\mhyphen 455}$ is statistically significant over a region of radius~1.2$^{\circ}$ around the position $\alpha$ = 08$^{\mathrm{h}}$ 35$^{\mathrm{m}}$ 00$^{\mathrm{s}}$, $\delta$ = -45$^{\circ}$ 36$^{\mathrm{\prime}}$ 00$^{\mathrm{\prime}\mathrm{\prime}}$ (J2000). The Vela~X region exhibits almost uniform $\gamma$-ray spectra over its full extent: the differential energy spectrum can be described by a power-law function with a hard spectral index $\Gamma$ = 1.32 $\pm$ 0.06$_{\mathrm{stat}}$ $\pm$ 0.12$_{\mathrm{sys}}$ and an exponential cutoff at an energy of (14.0~$\pm$~1.6$_{\mathrm{stat}}$~$\pm$~2.6$_{\mathrm{sys}}$) TeV. Compared to the previous H.E.S.S. observations of Vela~X the new analysis confirms the general spatial overlap of the bulk of the VHE $\gamma$-ray emission with the X-ray cocoon, while its extent and morphology appear more consistent with the (more extended) radio emission, contradicting the simple correspondence between VHE $\gamma$-ray and X-ray emissions. 
Morphological and spectral results challenge the interpretation of the origin of $\gamma$-ray emission in the GeV and TeV ranges in the framework of current models.}
{}

   \keywords{Radiation mechanisms: non-thermal -- Gamma rays: general-- ISM: individual objects: Vela~X -- ISM: individual objects: HESS J0835-455}
\authorrunning{H.E.S.S. collaboration}
%\titlerunning{Extent of the non-thermal emission from Vela~X at TeV energies} 
\titlerunning{Probing the extent of the emission from Vela~X at TeV energies}

   \maketitle

\section{Introduction}
Very-High-Energy ($E \ge$ 100 GeV) photon emission originating from about one hundred astrophysical sources populating the Galactic Plane has been detected in recent years. A significant fraction of Galactic VHE $\gamma$-ray sources are pulsar wind nebulae (PWNe), which are interpreted as bubbles filled with a plasma of highly energetic particles. PWNe are associated with pulsars, whose rotational energy is converted into the kinetic energy of a plasma of charged particles (see \citealp*{gaensler} and references therein for a recent review).

Vela~X was originally discovered by \cite{risbeth} as one
of three radio sources in the western part of the Vela constellation. \cite{risbeth} already presumed that Vela~X stands out in comparison
to the other two objects (Vela~Y, Vela~Z), as its radio spectrum
tends to be flatter. A review of radio observations of the region
was given by \cite{alvarez}. The size of the Vela~X radio emission
is coarsely given as 3$^{\circ}$ in the east-west direction and 2$^{\circ}$
in the north-south direction. Vela~X was interpreted as a PWN associated with the pulsar PSR~B0833-45 (\citealp{weipana}). A pulsar distance of about
290 pc is derived from radio and optical parallax measurements (\citealp{caraveo}; \citealp{dodson}). As a side-product of the parallax measurement, a proper motion velocity of 65 km s$^{-1}$ was found, which represents a comparatively slow motion with respect to typical pulsar velocities, that has prevented the pulsar from escaping from the nebula bubble. The Vela pulsar exhibits a rotational period of 89.3 ms (\citealp{large}) and a period derivative of 1.25 $\times$ 10$^{-13}$ s s$^{-1}$ (\citealp{lyne}), corresponding to a spin-down luminosity of 6.9 $\times$ 10$^{36}$ erg s$^{-1}$. 
PSR B0833-45 is the only middle-aged pulsar with a PWN for which a braking index has been measured ($n$ = 1.4, \citealp{lyne}). The commonly quoted characteristic age of the pulsar is 11 kyr (assuming $n$ = 3, that is, magnetic dipole radiation to be responsible for the spin-down of the pulsar).
A large step forward in the understanding of Vela~X came with the X-ray mission ${\it ROSAT}$ in the mid 1990s.
According to the ${\it ROSAT}$ data, the shell of the Vela supernova
remnant (SNR) is extended over a large region of diameter
$\approx8^{\circ}$. Beyond the shell several fragments have been found
with trajectories pointing back toward the pulsar, so that shell and pulsar
are associated as remnants of the same SN explosion (\citealp{aschenbach1}). An elongated X-ray filament was observed at the centre of Vela~X using ${\it ROSAT}$ (\citealp{markwardt}). The filament, primarily seen at energies between 1.3~keV and 2.4~keV, is oriented along the north-south direction. These measurements were confirmed by observations performed using ${\it ASCA}$ (\citealp{markwardt2}), which yielded an estimated length for this filament of about 45$'$ (from the pulsar to the centre of Vela~X).
The term {\it cocoon} stems from the original and nowadays obsolete interpretation of the X-ray filament as a jet outgoing from the pulsar and filling a cocoon about its head.
More detailed X-ray observations took place recently, covering a larger region including to the north of the cocoon: a hard X-ray emission has been detected just outside Vela~X in the energy range 2-10 keV using ${\it Suzaku/XIS}$ (\citealp{katsuda}). Observations performed with the ${\it Chandra}$ satellite have resolved a structured PWN within $\simeq$1$'$ of the pulsar, showing jet-like emission aligned with the direction of the pulsar's proper motion. These data have provided the prime evidence for the pulsar wind (\citealp{helfand}; \citealp{pavlova}; \citealp{pavlovb}). The X-ray flux in the 3-10 keV energy range measured with XMM-Newton/MOS and BeppoSAX/MECS shows a spectral index gradually softening from an index of 1.60 to 1.90 in annular integration regions within 15$'$ from the pulsar position (\citealp{mangano}).

Furthermore, extended emission above 18 keV from the Vela nebula on the northern side of the cocoon was observed with the ${\it INTEGRAL}$ satellite (\citealp{mattana}). This new diffuse component has no known counterparts and appears larger and more significant than the southern part, which is partially coincident with the soft X-ray cocoon. The origin of the northern hard X-ray emission might be due to the injection of relatively fresh particles into the nebula. The modelling of the spectrum within 6$^{\mathrm{\prime}}$ from the pulsar and in the energy range from 18 to 400 keV implies a magnetic field higher than 10 $\mu$G.

The Vela~X cocoon was observed with the ${\it VLA}$, revealing
a bright radio filament coincident with the eastern edge of the
X-ray filament (\citealp{frail}). Moreover, a network of similar
radio filaments was discovered across the whole radio PWN with
extended observations by the {\it Molonglo Observatory Synthesis Telescope} ({\it MOST}, \citealp{bock}), whereas only the central one is visible in X-rays. A suggested explanation for these filaments is that they arise from Rayleigh-Taylor (RT) instabilities, due to the accelerated expansion of the nebula into the surrounding medium. Such acceleration is expected to occur in the free expansion phase (for PWNe with an age of about 1 kyr), or in the reverberation phase, when the nebula is first adiabatically compressed by the reverse shock of the SNR and then re-expands due to the rising pressure inside the PWN. \cite{gelfand} presented the simulated evolution of a PWN inside a SNR over 100 kyr. Following their model, PWNe with ages of about 20 kyr are characterised by a large magnetisation parameter (i.e. the ratio of magnetic field to particle energy), and will show RT-filament growth. The RT instabilities will allow the magnetic field lines to rearrange themselves parallel to the filaments, so that the plasma particles are bound within this region. The concept of an aligned magnetic field is supported by the measurement of the orientation of the magnetic field by \cite{milne}.

The Vela pulsar is located at the northern edge of the radio nebula. The offset between the pulsar and the centre of the PWN can be explained by the interaction of the PWN with the SNR shell. While the forward shock (FS) of the SNR expands
into the ISM, the reverse shock (RS), after moving outwards behind the FS for the first several hundred years after the supernova explosion, reverts its direction and moves inwards towards the centre of the SNR. It eventually collides with the PWN developing inside the SNR, causing a compression and potentially a displacement of the nebula. A non-uniform density of the interstellar medium may affect the uniformity of the expansion velocity of the FS, and hence the reversal of the RS, in the sense that the RS will return earlier from a direction where the shock encounters a denser medium. The cocoon is ascribed to an asymmetric reverse shock, which stripped the bulk of the particles from around the pulsar, thus leaving behind a relic PWN developed along the main X-ray axis, which is perpendicular to the direction of the Vela pulsar proper motion (\citealp{blondin}). In the case of the Vela SNR, evidence for a non-uniform ISM density was indeed found: an atomic hydrogen density of 1-2 particles per cm$^3$ for the northern parts of the SNR (\citealp{dubner}), compared with a density of 0.1 particles per cm$^3$ for the southern ISM (\citealp{ferreira}). Thus, the RS is assumed to have returned first from the northern part of the SNR and to have pushed the PWN about 1$^{\circ}$ to the south, where the position of the nebula was eventually confined by the southern part of the RS.

The H.E.S.S. collaboration reported VHE $\gamma$-ray emission from Vela~X (\citealp{velaHESS}) and found its morphological extent to be coincident with the X-ray cocoon detected with ${\it ROSAT}$ to the south of the pulsar position. A differential energy spectrum was extracted with an integration radius of 0.8$^{\circ}$ around the formal centre at RA = 08$^{\mathrm{h}}$ 35$^{\mathrm{m}}$ 00$^{\mathrm{s}}$, Dec = -45$^{\circ}$ 36$^{\mathrm{\prime}}$ 00$^{\mathrm{\prime}\mathrm{\prime}}$ (J2000), within the energy range $\mathrm{0.55\mhyphen 65}$ TeV. A fit of a power law yielded a spectral index $\Gamma$ = 1.45 $\pm$ 0.09$_{\mathrm{stat}}$ $\pm$ 0.2$_{\mathrm{sys}}$ and an exponential cutoff at an energy of (13.8 $\pm$ 2.3$_{\mathrm{stat}}$ $\pm$ 4.1$_{\mathrm{sys}}$)~TeV. The total $\gamma$-ray flux, integrated in energy above 1~TeV, is (12.8 $\pm$ 1.7$_{\mathrm{stat}}$ $\pm$ 3.8$_{\mathrm{sys}}$)~$\times$~$10^{\mathrm{-12}}\,\mathrm{cm}^{\mathrm{-2}}\,\mathrm{s}^{\mathrm{-1}}$. The integration region will be referred to in the following as the inner test region. The formal centre will hereafter be called the central position.

Following the H.E.S.S. observations, \cite{dejager2} presented a two-zone leptonic model for Vela~X which assumes constant magnetic field and plasma density throughout the Vela~X region. In the model the Vela~X PWN consists of two emission regions; while the X-ray cocoon and the VHE $\gamma$-ray observations appear to define a central region, radio observations reveal a more extended area 2$^{\circ}$ $\times$ 3$^{\circ}$ in size. 
The model proposes two distinct electron spectra, responsible for the radio and X-ray/VHE $\gamma$-ray emission, respectively. For a magnetic field strength of 5 $\mu$G this model reproduces an Inverse Compton (IC) energy maximum in the multi-TeV domain as observed with H.E.S.S., while predicting the total hard X-ray cocoon emission above 20 keV measured using instruments onboard the ${\it INTEGRAL}$ satellite within a factor of two. De Jager et al. (2008a) also propose that a higher field strength in the radio filaments would imply fewer leptons in Vela~X and hence a fainter signal in the GeV energy domain.

The measurements of the spectral energy distribution (SED) were recently supplemented in the high-energy domain, from 100 MeV to 20 GeV, by observations made using the space-borne missions ${\it AGILE}$ (\citealp{pelli}) and ${\it Fermi}$ (\citealp{fermivela}). The spectral results of these measurements agree with the expectations of the scenario of \cite{dejager2} in which two particle populations are invoked to explain the high energy emissions in the GeV and TeV domains respectively. 
A recent attempt to unify these two populations and explain the steep spectral index measured with the {\it Fermi}-LAT (2.41 $\pm$ 0.09$_{\mathrm{stat}}$ $\pm$ 0.15$_{\mathrm{sys}}$ between 200 MeV and 20 GeV) as a consequence of particle escape has been proposed by \cite{hinton}. 

Alternative scenarios to interpret the Vela~X SED are also proposed (\citealp{berka}, \citealp{horns}), where both leptons
and hadrons contribute to the non-thermal emission, and
in particular the VHE emission would result mainly from the
hadronic contribution. 

This work reports on re-observations of VHE $\gamma$-ray
emission from the Vela~X PWN with the H.E.S.S. telescope array.
After a summary of the follow-up observations lately made with H.E.S.S., and a description of the data analysis applied to the investigation of Vela~X (section~\ref{sec:obs}), the results regarding the spectrum (section~\ref{sec:spec}), the surface brightness (section~\ref{sec:sb}) and the morphology (section~\ref{sec:morp}) of the VHE $\gamma$-ray emission are detailed. A discussion (section~\ref{sec:disc}) on the results and possible multi-wavelength interpretation concludes this work.

\section{H.E.S.S. observations and data analysis}
\label{sec:obs}
The H.E.S.S. instrument consists of four 13 m diameter Imaging Air Cherenkov Telescopes (IACTs) located in the Khomas Highland in Namibia, 1800 m above sea level, and began operation in 2003 (\citealp{berno}, \citealp{funk}). The H.E.S.S. working principle is based on the detection of faint Cherenkov light from the $\gamma$-ray-induced air showers in the atmosphere, above an energy threshold of about 100 GeV up to several tens of TeV. The four telescopes are operated in a coincidence mode, in which at least two of them must have been triggered for each event within a coincidence window of 60 ns. Each telescope has a mirror area of 107~m$^2$ and is equipped with a camera containing 960 photomultipliers. 
H.E.S.S. is ideally suited for the morphological study of extended VHE $\gamma$-ray sources due to its high sensitivity, angular resolution of a few arc minutes and large (5$^{\circ}$) field of view. This said, the Vela~X region is experimentally challenging: it is the most extended VHE $\gamma$-ray source detected so far and has a relatively low surface brightness.

\subsection{Observations and data selection}

A background for the measurement of VHE $\gamma$-rays 
arises from charged cosmic-ray particles, which also generate air
showers. Typical methods to estimate the background
make use of regions free of $\gamma$-ray sources in the same field of view. These off-target regions of typically circular or annular shape cover the same area as the on-target region and are placed such that on- and off-target regions represent reflections with respect to the pointing position (see \citealp{berge} and \citealp{aha06b} for more details). However, these (standard) methods are not ideally suited to observations of very extended sources such as Vela~X. Therefore, the analysis is performed with the so-called On/Off background
method (\citealp{weekes}) which estimates the background
from separate observations pointing away from Vela~X, and towards $\gamma$-ray source-free regions with comparable observation conditions (\citealp{west1}).
\begin{figure}[!t]
  \centering
\includegraphics[width=1.0\linewidth]{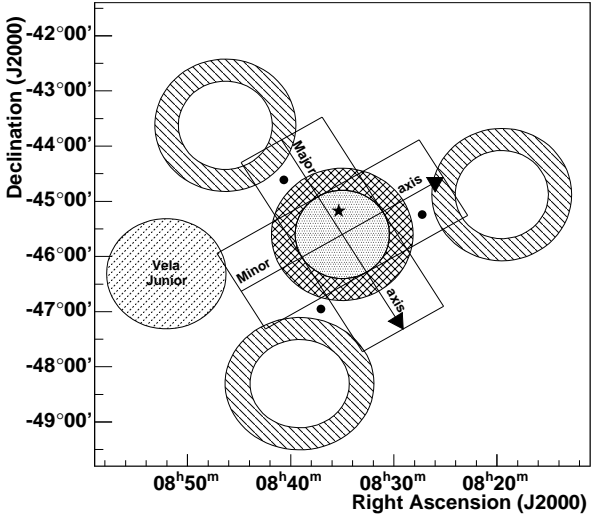}
  \caption{Schematic overview of the region around Vela~X.
The star marks the position of the Vela pulsar PSR B0833-45. The light-grey area indicates the
region 0.8$^{\circ}$ around the centre of the VHE $\gamma$-ray emission. The surrounding
cross-hatched ring has an outer radius of 1.2$^{\circ}$. The black dots mark
the pointing positions that allow background estimation in reflected
mode and which were chosen for the spectral analysis of the Vela~X outer region (the cross-hatched ring). The line-filled rings are the associated reflected off-source regions.
The two boxes are used to extract flux profiles (which are shown in Figure~\ref{slices}), where the arrows indicate the direction of the slices (as explained in section~\ref{sec:morp}). The position of the nearby SNR Vela Junior (\citealp{aha07}) is also shown.}
\label{overviewmap}
 \end{figure}
The Vela~X region was observed during several observation campaigns between 2004 and 2009. In 2004 data were taken using a method in which the source was offset by a small angular distance from the centre of the field of view ({\it wobble} mode). These observations were made at an offset of 0.5$^{\circ}$ in declination from the position of the Vela pulsar. Further observations from 2005 to 2007 were made surrounding the position measured as the centre of gravity of the VHE $\gamma$-ray excess. These were also taken in wobble mode, with a mean offset of 0.5$^{\circ}$. Data taken during the 2008 and 2009 observation campaigns enlarged the data set by 30\%. For the latter observations three more distant pointing positions offset by 1.3$^{\circ}$ from the central position were used, in order to study emission at a distance of up to 1.2$^{\circ}$ using the reflected annular background regions in the same field of view.
The annular background method estimates the cosmic-ray background contribution at each trial position on the sky, by integrating events in an annulus centred on that position, excluding potential source regions. These pointing positions are given as black dots in the overview image (Figure~\ref{overviewmap}), where the associated reflected off-source ring regions are also illustrated. In total 126 runs (each of 28 minutes observing time) were obtained after standard data quality selection (\citealp{aha06b}) and by requiring that all four telescopes were operative (to guarantee the best possible angular resolution). These runs correspond to a total observation time of 53.1 hours (12 hours of which are part of the total observation time analysed in the past and published in 2006), of which 14.8 hours were taken at a distance of 1.3$^{\circ}$ from the central position.
 
The On/Off background subtraction method requires particular care in the choice of the Off runs, which are taken from the general set of H.E.S.S. observations. Only those runs that can be considered free of VHE $\gamma$-ray sources are taken as Off runs. Each of the 126 On runs is paired with an Off run taken under matching observation conditions. After applying the aforementioned quality cuts to a list of potential Off runs, the selection is achieved by pairing On and Off runs such that the difference in the values of three relevant quantities is minimised. In order of their impact on the background rate, these quantities are: 1) The zenith angle of the observation: Vela~X was observed at different altitudes in the sky (zenith angles ranging from 20$^{\circ}$ to 45$^{\circ}$), which correspond to different energy thresholds. 2) The muon detection efficiency, which is used (\citealp{bolz}) to take into account the changing optical efficiency of the telescopes: Mainly due to the ageing of the mirrors, the muon detection efficiency decreased by more than 20\% over the course of the Vela~X observation campaigns (2003-2009). 3) The trigger rate: This is affected by variable atmospheric conditions, in particular clouds and aerosol formations.

Residual differences between on-target and background runs are a potential source of systematic uncertainties. Therefore, in order to assess these systematics, parts of the observations were performed pointing 1.3$^{\circ}$ off the central position (as mentioned above) to provide data suitable for analysis with the in-field background method.

\subsection{Reconstruction and data analysis}
\label{sec:recoda}

We employed the $X_\mathrm{eff}$ analysis (\citealp{dubois}) 
for the selection
of $\gamma$-ray events and for the suppression of cosmic-ray
background events. $X_\mathrm{eff}$ denotes a multivariate analysis
method developed to improve signal-to-background discrimination, which
is important in searches for weak signals and morphological studies of
extended sources (\citealp{xeff}). The $X_\mathrm{eff}$
method improves the separation of $\gamma$ and cosmic-ray events
compared to the standard H.E.S.S. analysis (\citealp{aha06b}), by exploiting the complementary discriminating
variables of three reconstruction methods\footnote{The three
  reconstruction methods are referred to as Hillas
  (\citealp{hillas}), Model (\citealp{model}) and 3D-model (\citealp{model3D}; \citealp{godo}).} in use in the H.E.S.S. analysis. The resulting unique discriminating variable $X_\mathrm{eff}$ acts as an event-by-event $\gamma$-misidentification probability estimator. The definition of the $X_\mathrm{eff}$ probability function follows the relation:
\begin{equation}
X_\mathrm{eff}(d_i) = \frac{{\eta} {\displaystyle\prod_{j}H_j(d_i)}}{(1-{\eta}){\displaystyle\prod_{j}G_j(d_i)+{\eta}\prod_{j}H_j(d_i)}},
\label{global}
\end{equation}
where $d_i$ are the discriminating variables of three reconstruction methods; $G_j(d_i)$ and $H_j(d_i)$ are the one-dimensional {\it probability density functions} ({\it p.d.f.}s) for events identified as $\gamma$-ray-like ($G$) and hadron-like ($H$) (the product of individual {\it p.d.f.}s replaces the global multi-dimensional ones since the variables $d_i$ are highly uncorrelated); $\eta$ is the misidentified fraction of the $gamma$ class of events (i.e. the relative background fraction). The final $\gamma$-ray event selection was achieved with the set of cuts adapted to the detection of faint sources ($\eta=0.5$, $X_\mathrm{eff,cut}=0.3$) and a cut in the reconstructed image charge requiring more than 80 photo-electrons.
With this approach the background rejection is improved over the full energy range of H.E.S.S. allowing for an increased quality factor (defined as the ratio of $\gamma$-ray detection efficiency to the square root of residual cosmic-ray-proton efficiency) ranging from a factor of 1.6 (at 0.3 TeV) to 2.8 (at E$>$10 TeV) with respect to the standard Hillas reconstruction method. \\
Similarly, the shower direction and energy of $\gamma$-ray events are
reconstructed by means of an  estimator composed of the
directions and energies reconstructed from the three methods
previously referred to (\citealp{lyons}; \citealp{xeff}). A weighting factor taking into account the
covariance matrices between estimates and minimising the corresponding
variance is applied to the individual reconstruction variables
($z_i$), to provide the best unbiased estimate of the observable
$\hat{z}$. This procedure improves both the energy resolution ($\simeq$10\% at E=1 TeV and $\simeq$15-20\% for E$>$10 TeV) and the angular resolution, resulting in a point spread function (PSF; 68\% containment radius) between 0.05$^{\circ}$ and 0.07$^{\circ}$. 

For cross-check purposes the data analysis was also conducted, within
the same analysis software, by applying each of the three above
mentioned reconstruction methods individually. Furthermore, an
independent standard Hillas analysis (i.e. using an independent data
calibration, instrument response functions and analysis software) was
performed (\citealp{gluck}). In this independent analysis
the event reconstruction is based on the {\it scaled widths} method
using {\it standard cuts} as documented in \cite{aha06b} and \cite{aha06c}. All analyses
yielded compatible flux measurements within a 10\% systematic error and
a variation in spectral index of about 4\%.

All analyses used in this work use separate observation runs for the background rejection to the signal collection (On/Off observation method). This could imply a residual systematic shift in the estimated background rate due to the differences in the observation conditions between the On and Off fields.
These potential systematic shifts are considered by an acceptance correction factor applied to the absolute normalisation $\alpha$ which is defined as the ratio between the live-times of the On and Off data. 

Several tests were carried out to validate the systematic uncertainties and to verify the stability of the spectral results. 
For the subsets of data (for which it was possible) the background rates resulting from the application of the On/Off method and the reflected background regions method were compared. An approximate 8\% uncertainty on the flux normalisation and a corresponding systematic uncertainty of 0.08 on the spectral index were estimated. These values were confirmed by applying the On/Off analysis to other $\gamma$-ray sources as well as to dark-field data sets (i.e. sky test regions without any detected VHE source). Performing the On/Off analysis independently on three distinct and statistically equivalent sub-sets of data enabled us to quantify 10\% and 0.05 sytematic errors on the flux and the spectral index, respectively. 
Results obtained for all of the analysis chains with different sets of selection-cuts (namely {\it hard}, {\it standard} and {\it loose} cuts as explained in \cite{aha06b} and \cite{aha06c}) lead to an 8\% error on the flux and to a systematic uncertainty of 0.06 on the spectral index, while these errors are 10\% and 0.05, respectively, when the results of the two independent analysis chains are compared. \\
A summary of various estimated contributions to the systematic errors on the results of the spectral analysis is given in Table~\ref{T:cuts}.

\begin{table}[!ht]
\caption{Various estimated contributions to the systematic error of the obtained flux and spectral index for this specific analysis. Common H.E.S.S. analysis uncertainty contributions studied in \cite{aha06b} are also quoted ($^{\dag}$). 
\label{T:cuts}}
\centering
\begin{tabular}{c c c}
\hline
\hline
Uncertainty          &                 Flux          &     Index\\
\hline
\hline
Monte Carlo shower interactions$^{\dag}$ &        1\% &\\

Monte Carlo atmospheric simulation$^{\dag}$    &      10\% &\\

Broken pixels$^{\dag}$     &             5\%&\\

Live time$^{\dag}$        &                1\%&\\

Background estimation             &          8\%&                 0.08\\
Variability as a function of data sets &      10\%                   &    0.05\\
Analysis selection cuts &             8\% &                0.06\\
Difference between analysis chains &             10\% &                0.05\\
\hline
Total                          &             21\%       &          0.12\\
\hline
\end{tabular}
\end{table}

\section{Results}

The analysis yields an excess of 4010 $\gamma$-ray events in the
Vela~X region, within a radial distance of 1.2$^{\circ}$ around the
central position $\alpha$ = 08$^{\mathrm{h}}$ 35$^{\mathrm{m}}$
00$^{\mathrm{s}}$, $\delta$ = -45$^{\circ}$ 36$^{\mathrm{\prime}}$
00$^{\mathrm{\prime}\mathrm{\prime}}$ (J2000) and applying the minimum
threshold of 0.75 TeV on the event reconstructed energy. The nominal
energy threshold for observations with the H.E.S.S. experiment under
ideal conditions and for a source observed at about 20$^{\circ}$
zenith angle is $\sim$ 0.2 TeV. This threshold deteriorated to
$\sim$~0.4 TeV due to the degrading optical efficiency of the mirrors
associated with ageing. As mentioned, Vela~X was observed at different altitudes in the sky with effective energy thresholds ranging from 0.4 TeV to more than 0.7 TeV. An energy cut $E_\mathrm{th}$=0.75 TeV was chosen to provide a homogeneous event sample with respect to the different threshold energies. The $\gamma$-ray excess has a significance of 27.9 $\sigma$, while the signal-to-background ratio is 0.6. 
In the ring between the radii 0.8$^{\circ}$ and 1.2$^{\circ}$ around the central position a pre-trials significance of 7.5 $\sigma$ is measured.   
When the reflected background model is applied to the dedicated data sub-set (sample of observations performed with 1.3$^{\circ}$ off-pointing) a significance of 5.3 $\sigma$ is found. The difference in the significances for the two methods of background subtraction reflects the differences in observation times between On/Off data and data analysed with the reflected background method. Table~\ref{stati} shows the summary of the event statistics.

\begin{table}[!b]
  \caption{Number $N_\mathrm{on}$ of events in the test regions, the $\alpha$ factors,
the number $N_\mathrm{b}$ of background events (corresponding to $N_\mathrm{off}\times\alpha$), the resulting significance $\sigma$ and the number of excess events $N_\mathrm{\gamma}$ for the On/Off background model. The numbers are obtained by applying a minimum energy threshold $E_\mathrm{th}$ of 0.75 TeV, while in the last line of the table the total numbers without any minimum cut on the energy are given.}
  \label{stati}
  \centering
  \begin{tabular}{ccccccc}
  \hline
  \hline
   Region &$E_\mathrm{th}$ ($\mathrm{TeV}$) &$N_\mathrm{on}$& $\alpha$&$N_\mathrm{b}$&$\sigma$&$N_\mathrm{\gamma}$ \\
   \hline
   \hline
   Inner& 0.75 & 6870 & 1.14 & 3591.1 & 31.1 & 3278.9   \\
   Ring& 0.75 & 4763 & 1.14 & 4031.8 & 7.5 & 731.2   \\
   Total& 0.75 & 11633 & 1.14  & 7623 & 27.9 & 4009.9   \\
  \hline
   Total&  & 45811 & 1.07 & 39822.6 & 20.1 & 5988.2    \\
  \hline
   \end{tabular}
  \end{table}

\subsection{Spectral analysis}
\label{sec:spec}

The spectral analysis of the VHE $\gamma$-ray emission was conducted for the $\gamma$-like events reconstructed with energies higher than $E_\mathrm{th}$=0.75 TeV. All differential energy spectra of the VHE $\gamma$-ray emission have been obtained using a forward-folding method (\citealp{piron}) based on the measured energy-dependent on-source and off-source distributions.

\begin{table*}[!t]
  \caption{Best fit results of the differential energy spectra extracted for the inner and the external ring test regions as well as integrating over the total Vela~X extension. The best approximation corresponds to an exponentially cut-off power-law function 
($\mathrm{d} \Phi(E) / \mathrm{d} E$ = $N_\mathrm{0}(E/\mathrm{1 TeV})^{-{\it \Gamma}}e^{-E/E_\mathrm{cut}}$).}
  \label{tabspe}
  \centering
  \begin{tabular}{c c c c c}
  \hline
  \hline
   Region & ${\it \Gamma}$ & $E_\mathrm{cut}$ ($\mathrm{TeV}$) & $N_\mathrm{0}$ ($10^{\mathrm{-12}}\,\mathrm{cm}^{\mathrm{-2}}\,\mathrm{s}^{\mathrm{-1}}\,\mathrm{T
eV}^{\mathrm{-1}}$) & $\Phi_{>1\mathrm{TeV}}$ ($10^{\mathrm{-12}}\,\mathrm{cm}^{\mathrm{-2}}\,\mathrm{s}^{\mathrm{-1}}$) \\
   \hline
   \hline
   Inner & 1.36 $\pm$ 0.06$_{\mathrm{stat}}$ $\pm$ 0.12$_{\mathrm{sys}}$ & 13.9 $\pm$ 1.6$_{\mathrm{stat}}$ $\pm$ 2.6$_{\mathrm{sys}}$ & 11.6 $\pm$ 0.6
$_{\mathrm{stat}}$ $\pm$ 2.4$_{\mathrm{sys}}$ & 16.0 $\pm$ 1.3$_{\mathrm{stat}}$ $\pm$ 3.3$_{\mathrm{sys}}$ \\

   Ring & 1.14 $\pm$ 0.2$_{\mathrm{stat}}$ $\pm$ 0.12$_{\mathrm{sys}}$ & 9.5 $\pm$ 2.7$_{\mathrm{stat}}$ $\pm$ 1.7$_{\mathrm{sys}}$ & 3.3 $\pm$ 0.6$_{\mathrm{stat}}$ $\pm$ 0.7$_{\mathrm{sys}}$ & 4.9 $\pm$ 1.4$_{\mathrm{stat}}$ $\pm$ 1.1$_{\mathrm{sys}}$ \\

   Total & 1.32 $\pm$ 0.06$_{\mathrm{stat}}$ $\pm$ 0.12$_{\mathrm{sys}}$ & 14.0 $\pm$ 1.6$_{\mathrm{stat}}$ $\pm$ 2.6$_{\mathrm{sys}}$ & 14.6 $\pm$ 0.8$
_{\mathrm{stat}}$ $\pm$ 3.0$_{\mathrm{sys}}$ & 21.0 $\pm$ 1.9$_{\mathrm{stat}}$ $\pm$ 4.4$_{\mathrm{sys}}$  \\
  \hline
   \end{tabular}
  \end{table*}

\begin{figure}[!t]
  \centering
  \includegraphics[width=1.0\linewidth]{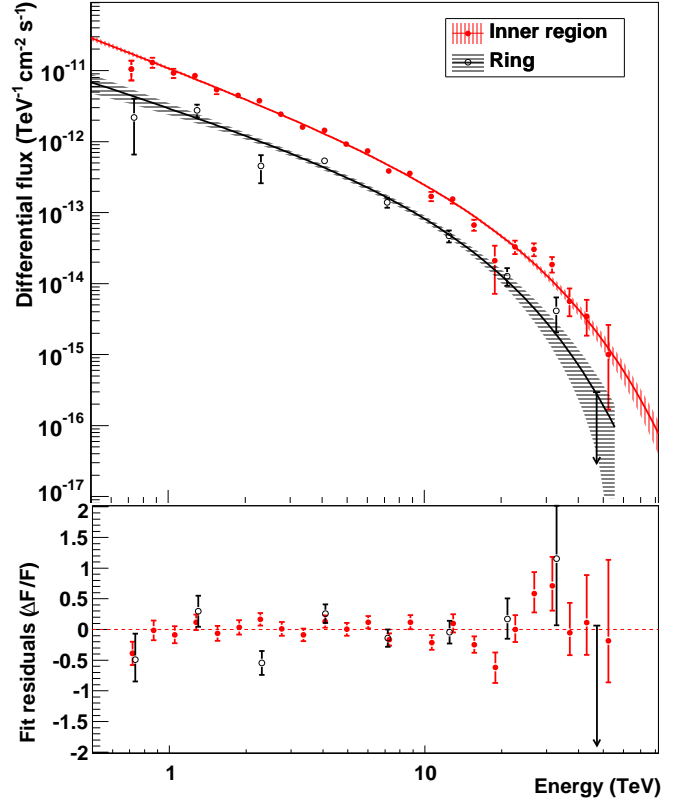}
  \caption{Differential $\gamma$-ray spectrum of Vela~X in the
TeV energy range. Filled red circles: inner integration region $<0.8^{\circ}$; open black circles: ring extension (between 0.8$^{\circ}$ and 1.2$^{\circ}$). Both spectra are fitted with a power law with exponential cutoff. The shaded bands correspond to the statistical uncertainty of the fit.}
  \label{spectreall}
\end{figure}
\begin{figure}[!b]
  \centering
  \includegraphics[width=0.9\linewidth]{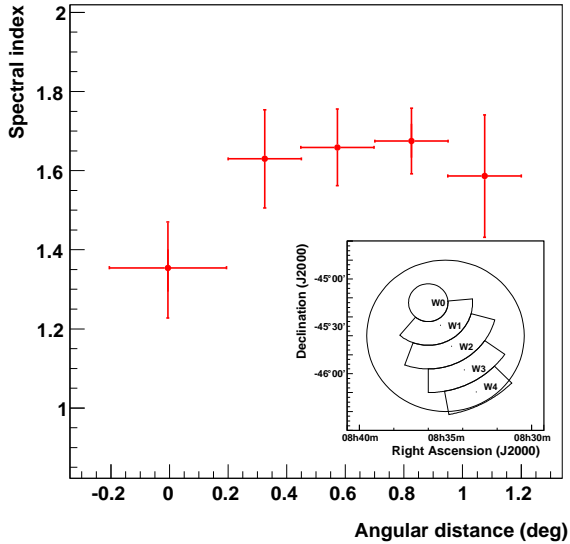}
  \caption{Spectral indices as a function of radial distance to the pulsar position along the major axis of Vela~X. The insert shows a graphic of the five sectors on the sky-map (with radial distances of 0$^{\circ}$ ($W0$), 0.325$^{\circ}$ ($W1$), 0.575$^{\circ}$ ($W2$), 0.825$^{\circ}$ ($W3$) and 1.075$^{\circ}$ ($W4$), respectively) from which the differential energy spectra were extracted.} 
  \label{spectrewi}
\end{figure}
A differential energy spectrum extracted from the inner region (within a radius $<$ 0.8$^{\circ}$) of Vela~X is shown in Figure~\ref{spectreall} in red. The best approximation of the energy spectrum corresponds to 
an exponentially cut-off power-law function ($\mathrm{d} \Phi(E) / \mathrm{d} E$ = $N_\mathrm{0}(E/\mathrm{1 TeV})^{-{\it \Gamma}}e^{-E/E_\mathrm{cut}}$) with an index ${\it \Gamma}$=1.36 $\pm$ 0.06$_{\mathrm{stat}}$ $\pm$ 0.12$_{\mathrm{sys}}$, a cutoff energy $E_\mathrm{cut}$=(13.9 $\pm$ 1.6$_{\mathrm{stat}}$ $\pm$ 2.6$_{\mathrm{sys}}$)~TeV and $N_\mathrm{0}$=(11.6 $\pm$ 0.6$_{\mathrm{stat}}$ $\pm$ 2.4$_{\mathrm{sys}}$)~$\times$~$10^{\mathrm{-12}}\,\mathrm{cm}^{\mathrm{-2}}\,\mathrm{s}^{\mathrm{-1}}\,\mathrm{TeV}^{\mathrm{-1}}$. 
The spectrum of the VHE $\gamma$-ray emission in the outer ring (between 0.8$^{\circ}$ and 1.2$^{\circ}$) is drawn in black in Figure~\ref{spectreall}. A photon index ${\it \Gamma}$=1.14 $\pm$ 0.2$_{\mathrm{stat}}$ $\pm$ 0.12$_{\mathrm{sys}}$, a cutoff energy $E_\mathrm{cut}$=(9.5 $\pm$ 2.7$_{\mathrm{stat}}$ $\pm$ 1.7$_{\mathrm{sys}}$)~TeV and $N_\mathrm{0}$=(3.28 $\pm$ 0.58$_{\mathrm{stat}}$ $\pm$ 0.7$_{\mathrm{sys}}$)~$\times$~$10^{\mathrm{-12}}\,\mathrm{cm}^{\mathrm{-2}}\,\mathrm{s}^{\mathrm{-1}}\,\mathrm{TeV}^{\mathrm{-1}}$ are obtained for an exponentially cut-off power law. 
%($\mathcal{D}$ = 2 $\mathrm{log}$ $(\mathcal{L}^{EC}/\mathcal{L}^{PL})$) 
Since the ring spectrum derived is consistent with the one of the inner region, the total spectral shape of Vela~X can be calculated from a 1.2$^{\circ}$ circular region around the central position. A fit to the spectrum of the entire Vela~X region yields a spectral index ${\it \Gamma}$=1.32 $\pm$ 0.06$_{\mathrm{stat}}$ $\pm$ 0.11$_{\mathrm{sys}}$, $E_\mathrm{cut}$=(14.0 $\pm$ 1.6$_{\mathrm{stat}}$ $\pm$ 2.6$_{\mathrm{sys}}$)~TeV and $N_\mathrm{0}$=(14.6 $\pm$ 0.82$_{\mathrm{stat}}$ $\pm$ 3.0$_{\mathrm{sys}}$)~$\times$~$10^{\mathrm{-12}}\,\mathrm{cm}^{\mathrm{-2}}\,\mathrm{s}^{\mathrm{-1}}\,\mathrm{TeV}^{\mathrm{-1}}$. A pure power-law fit is clearly disfavoured for all spectra by relatively high likelihood ratio values ($>$ 25) between the exponentially cut-off power-law and the power-law fit hypotheses. The results of the fits are summarised in Table~\ref{tabspe}.

%

%It has been shown that the X-ray spectral index in the 3-10 keV energy range is gradually softening from an index of 1.60 to 1.90 (respectively from 1.50 to 1.66 for a different choice of integration regions) near the pulsar compared to the surrounding regions (\citealp{mangano}). 
In order to probe the hypothesis of potential radiative cooling of the electrons responsible for the IC origin of VHE $\gamma$-rays during their propagation away from the pulsar, a spectral softening is searched for in the H.E.S.S. data. To this end, spectra were extracted from a circular region with a radius of 0.2$^{\circ}$ around the position of the pulsar as well as from four ring sectors centred on the pulsar position and lying along the main axis of the VHE PWN (from north-west to south-east) at a radial distance of 0.3$^{\circ}$, 0.6$^{\circ}$, 0.8$^{\circ}$ and 1.1$^{\circ}$, respectively, according to the hypothesis of advection of seed electrons. The regions (labelled $W0$ through $W4$; see Figure~\ref{spectrewi} and insert therein) were chosen such that equivalent values of signal significance are achieved.       
A power-law fit limited to the energy range 0.75 to 10 TeV, disregarding the energy cutoff (because of the limited statistics of events at higher energy), was performed for each sector. This study yields compatible spectral indices included within 1.58 $\pm$ 0.15 and 1.67 $\pm$ 0.08 for all sectors but the one around the pulsar ($W0$) whose spectrum exhibits a photon index 1.35 $\pm$ 0.12 (see Figure~\ref{spectrewi}). Although some variation of the spectral index is found, the effect is limited to the immediate vicinity of the pulsar and is not statistically significant.

 \begin{figure}[!t]
  \centering
 \includegraphics[width=1\linewidth]{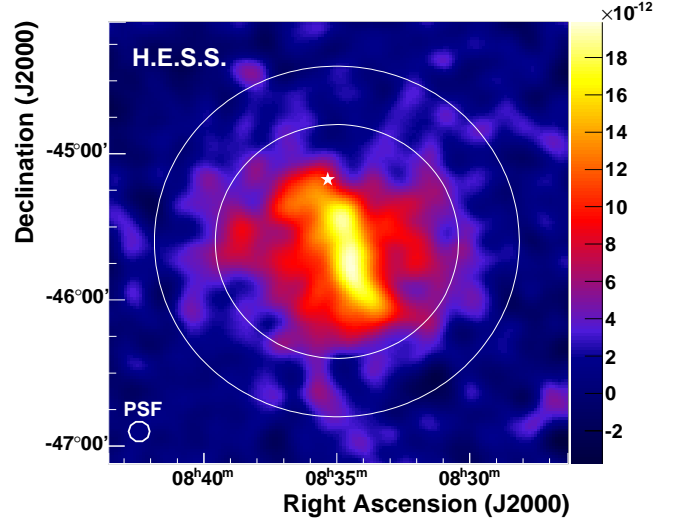}
  \caption{H.E.S.S. VHE $\gamma$-ray surface brightness ($\mathrm{cm}^{\mathrm{-2}}\,\mathrm{s}^{\mathrm{-1}}\,\mathrm{deg}^{\mathrm{-2}}$) of Vela~X integrated between 0.75 TeV and 70 TeV and 0.07$^{\circ}$ Gaussian smoothing width. The 0.07$^{\circ}$ PSF achieved with the $X_\mathrm{eff}$ method applied in this analysis is also shown for comparison. The circles are drawn with radii of 0.8$^{\circ}$ and 1.2$^{\circ}$, respectively, around the central position of the VHE $\gamma$-ray emission. The white star marks the position of the pulsar PSR B0833-45.} 
  \label{excessmap80}
 \end{figure}

\subsection{Surface brightness}
\label{sec:sb}
The surface brightness map of Vela~X and its surroundings shown in Figure~\ref{excessmap80} reveals one of the largest objects in the VHE $\gamma$-ray domain. The map was derived by comparing a measured
particle rate ($n$) with an assumed reference spectrum (${\mathrm dN}/{\mathrm dE}$). 
The best approximation of the differential energy spectrum, ${\mathrm dN}/{\mathrm dE}$ $\propto$ $E^{-{1.36}}e^{-E/(13.9\;\mathrm{TeV})}$ (resulting from the spectral analysis discussed in section~\ref{sec:spec}), has been assumed as reference spectrum in the calculation of the surface brightness map.
\begin{figure}[!ht]
  \centering
  \includegraphics[width=1.0\linewidth]{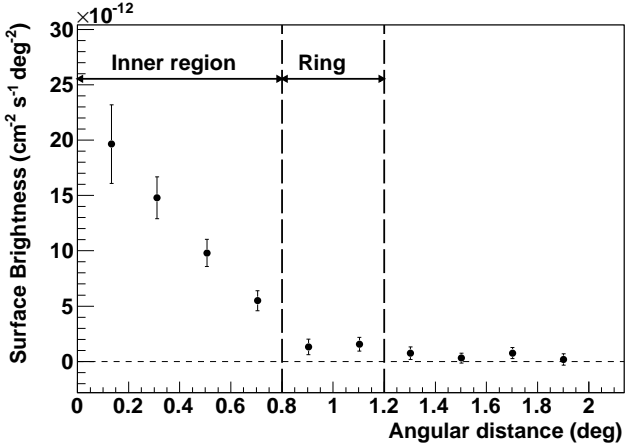}
  \caption{VHE $\gamma$-ray averaged surface brightness around the central position. The error bars reflect the statistical uncertainties. The vertical lines mark the 0.8$^{\circ}$ and 1.2$^{\circ}$ radii of the outer ring region, which contains significant emission.}
  \label{radidist}
\end{figure}
For each run, a number of expected counts can be defined and results from the integration over a given energy range of the reference spectrum multiplied by the effective area ($A_{x,y}$):
 \begin{equation}
G_{x,y}^\mathrm{run} = T ~ \int_{E_{\mathrm{th}}}^{\infty} A_{x,y}(E)~  \, \frac{\mathrm dN}{\mathrm dE}(E)~ \mathrm dE,
 \end{equation}
with $T$ being the observation time of the run and $E_{\mathrm{th}}$ the threshold energy appropriate for the specific observation (0.75 TeV commonly applied to all observation runs in this analysis). The normalization of the reference spectrum can be chosen arbitrarily since it will cancel out in the end. 
Several effects on the effective area $A_{x,y}(E,\theta,\rho,\epsilon)$ are taken into account through its dependency on: a)
the energy ($E$); b) the zenith angle ($\theta$) of the telescope pointing direction; c) the offset ($\rho$) of the $\gamma$-ray trajectory to the pointing; d) the optical efficiency ($\epsilon$) of the telescope reflectors. 

The number of expected $\gamma$-ray counts ($G_{x,y}$) in each cell of the surface map ($x,y$) results from the sum of expected counts over all runs:

 \begin{equation}
G_{x,y} = \sum\limits_\mathrm{runs}G_{x,y}^\mathrm{run}.
\end{equation}

Finally, the brightness $B_{x,y}$ per each bin ($x,y$) of the map is calculated as the ratio of
measured ($n_{x,y}$) and expected events ($G_{x,y}$), weighted with the integral of the reference spectrum and divided by the solid angle of the integration region ($s_{x,y}$) in the energy range where the spectral analysis is conducted ($E_\mathrm{th}$=0.75 TeV and $E_\mathrm{M}$=70 TeV):
 \begin{equation}
B_{x,y}  = \frac{1}{s_{x,y}} ~ \frac{n_{x,y}}{G_{x,y}} ~ \int_{E_\mathrm{th}}^{E_\mathrm{M}} \frac{\mathrm dN}{\mathrm dE}(E)~ \mathrm dE.
  \end{equation}
The radial profile of the surface brightness is presented in Figure~\ref{radidist}. It has been extracted from the surface brightness map by averaging over concentric rings around the central position. 
The VHE signal stretches out to a radius of 1.2$^{\circ}$. A residual excess which is not considered significant ($\leq$ 3~$\sigma$) is obtained over a larger radial extension (from 1.2$^{\circ}$ to 1.6$^{\circ}$).
Besides the strong signal in the inner region with a mean value of surface brightness of $\sim$~12~$\times$~10$^{-12}$~cm$^{-2}$~s$^{-1}$~deg$^{-2}$, there is a weaker signal spread over the ring extension with a mean $\gamma$-ray flux of $\sim$~1.4~$\times$~10$^{-12}$~cm$^{-2}$~s$^{-1}$~deg$^{-2}$. 

\begin{figure*}[!ht]
  \centering
 \includegraphics[width=0.45\linewidth]{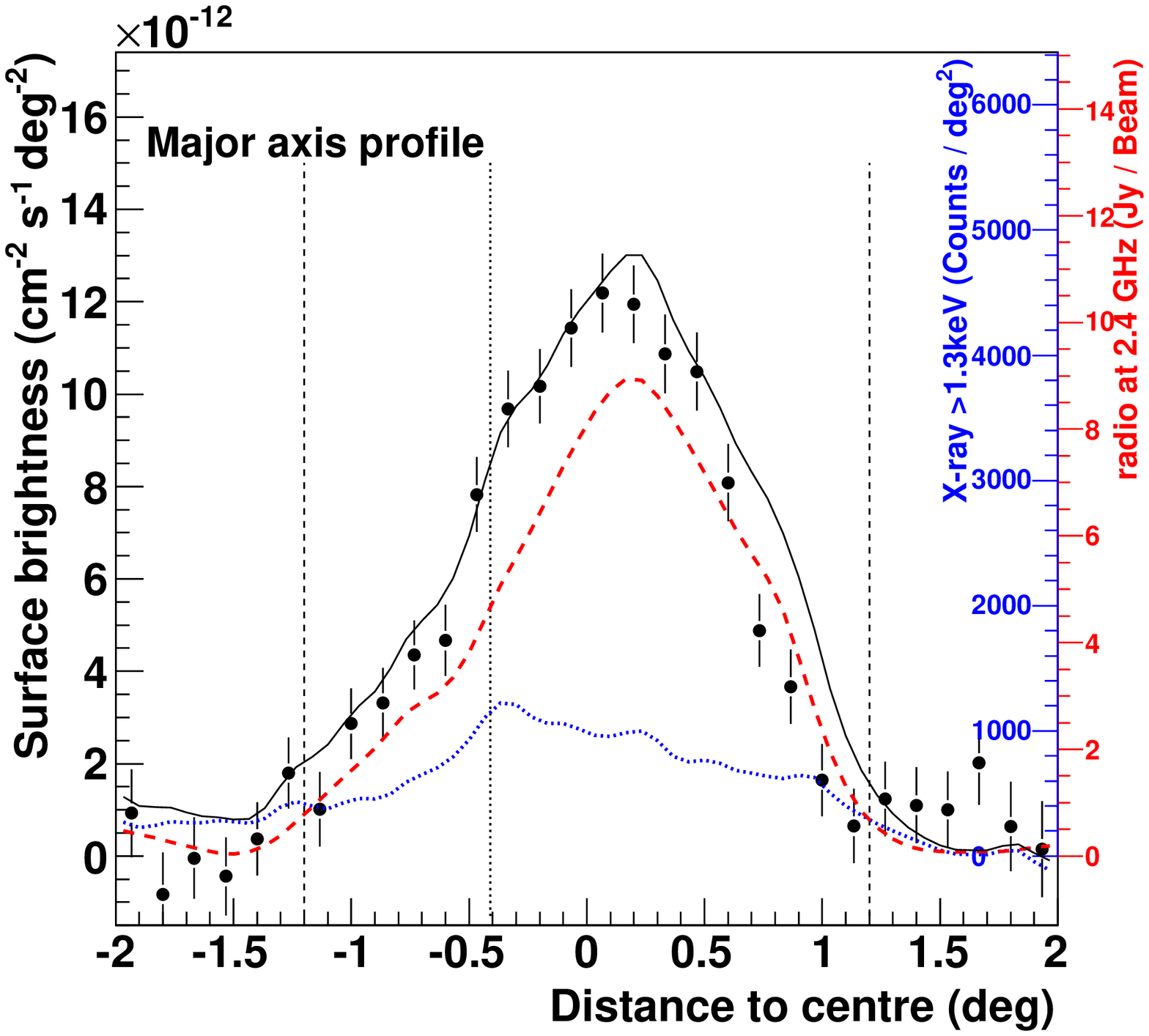}
  \includegraphics[width=0.45\linewidth]{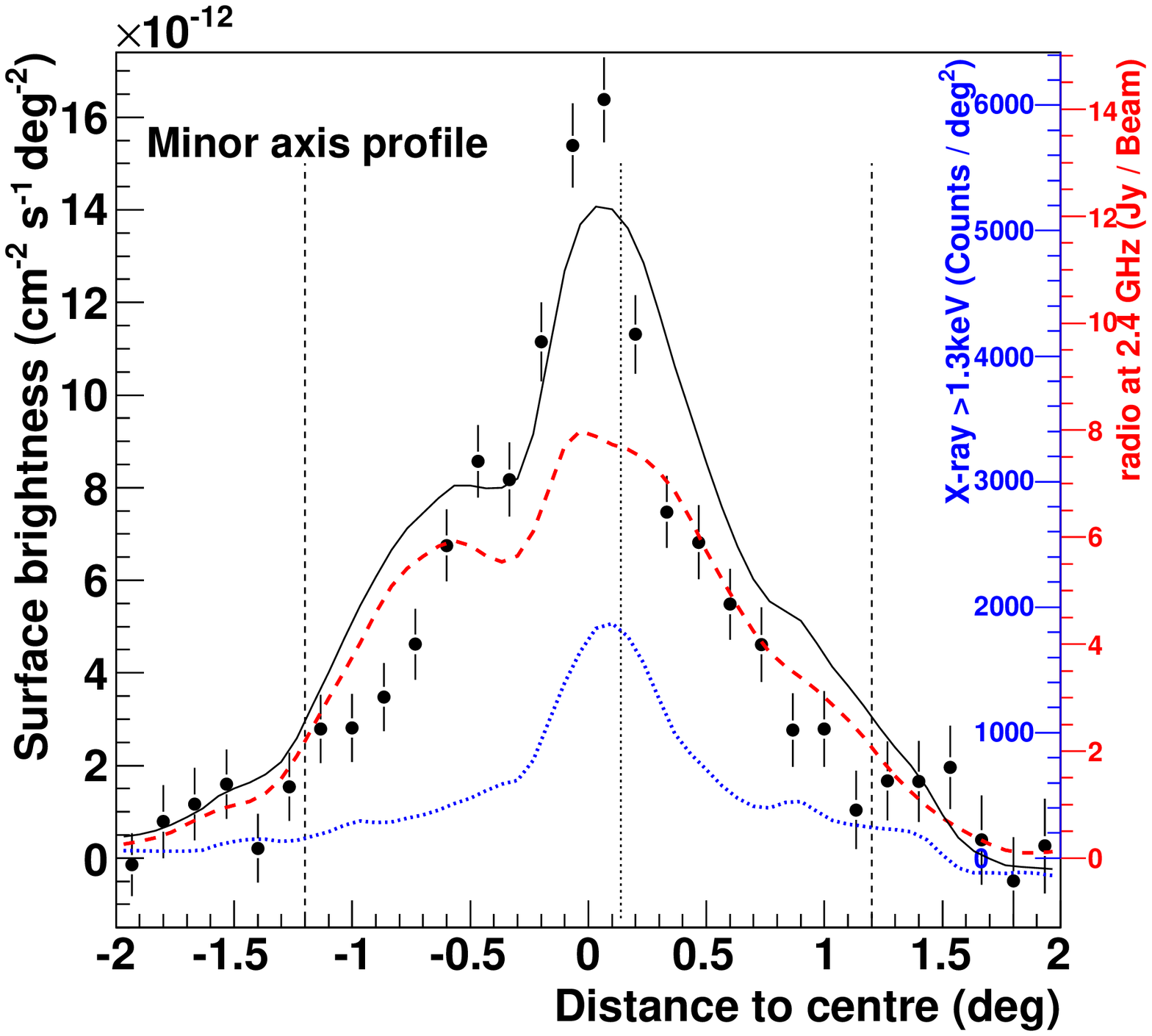}
  \caption{VHE $\gamma$-ray surface brightness profiles for E$>$0.75 TeV (filled black circles). The profiles through the central position are aligned parallel to the axis rotated 30$^{\circ}$ anticlockwise from the north, which is the major axis (left-hand panel, cf. Fig.1), and perpendicular to this direction, along the minor axis (right-hand panel). Profiles of the radio flux at 2.4 GHz (\citealp{duncan}) are drawn as red dashed lines, while X-ray data $>1.3$ keV (\citealp{aschenbach2}) and after removing the bright emission at the pulsar position, are drawn as blue dotted lines. These profiles are scaled by the normalisation factors resulting from the best fit of the linear combination (see equation~\ref{linear}) to the surface brightness map. Their sum, shown by black lines, is proposed as model for the VHE profiles. The corresponding absolute scales are shown on the right side of the plots in red for the radio flux and in blue for the X-ray data. The vertical lines mark the exterior borders of the ring extension (short-dashed lines) and the projected position of the pulsar within (dotted line).}
  \label{slices}
\end{figure*}

\subsection{Morphology}
\label{sec:morp}

The H.E.S.S. collaboration reported an intrinsic width of the VHE $\gamma$-ray flux profile from the Vela~X region, fitted by a Gaussian function, of 0.48$^{\circ}$ in south-west to north-east
direction and 0.36$^{\circ}$ in south-east to north-west direction
(\citealp{velaHESS}). In the following, this extent will be revisited using two perpendicular
profiles through the central position with a width of 1.6$^{\circ}$
and a length of 4.0$^{\circ}$ (see Figure~\ref{overviewmap}). The line oriented 30$^{\circ}$ anticlockwise
to north has been chosen as the axis for longitudinal
slices (the major axis as in Figure~\ref{overviewmap}). This axis is in better agreement with the orientation of
the spatial distribution of the VHE signal and is still close to the
axis derived in the previous analysis (\citealp{velaHESS}) so that the profiles can be compared. The filled black circles in Figure~\ref{slices} correspond to the VHE $\gamma$-ray surface brightness profiles along the two orthogonal axis.

\begin{figure*}[!ht]
  \centering
\hspace{-6mm}
 \includegraphics[width=0.5\linewidth]{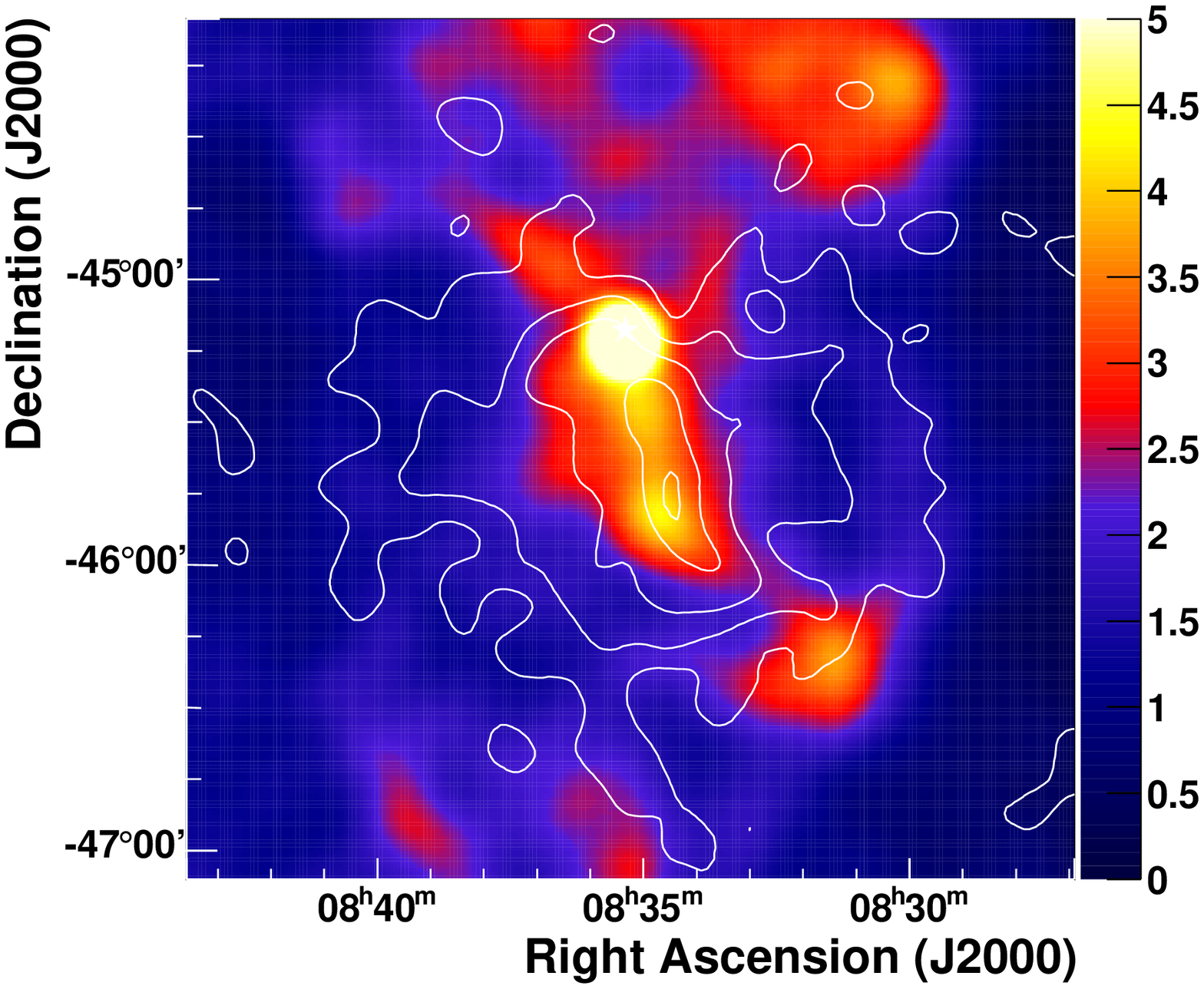}
\hspace{-6mm}
  \includegraphics[width=0.5\linewidth]{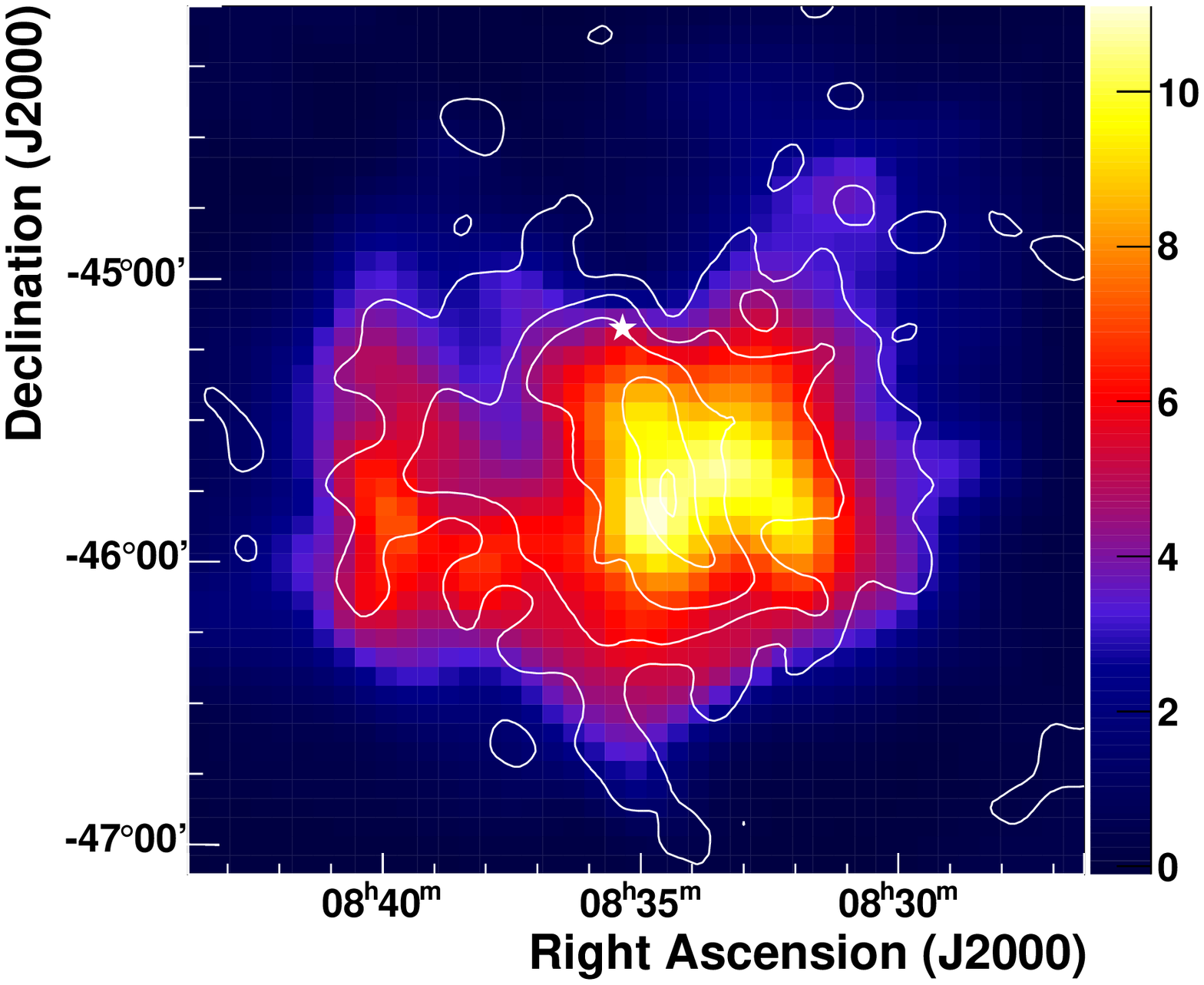}
  \caption{Left-hand panel: X-ray kcounts/deg$^2$ sky-map for energies E$>$1.3~keV measured with ${\it ROSAT}$ as from \cite{aschenbach2} and Gaussian-smoothed with a radius of 0.07$^{\circ}$. Note that the colour scale truncates the strong emission close to the pulsar at a level of 5 kcounts/deg$^2$. Right-hand panel: Radio sky-map at 2.4 GHz (Jy/Beam and half-power beam width of 0.17$^{\circ}$ measured) with the ${\it Parkes}$ telescopes as from \cite{duncan}. White contours correspond to VHE $\gamma$-ray surface brightness of 0.3, 0.6, 1, 1.6 and 1.9 $\times$ $10^{\mathrm{-11}}\,\mathrm{cm}^{\mathrm{-2}}\,\mathrm{s}^{\mathrm{-1}}\,\mathrm{deg}^{\mathrm{-2}}$.}
  \label{3wl}
\end{figure*}

The shape of the VHE $\gamma$-ray profiles along the major axis is well fit by a Gaussian function. The resulting width of (0.52~$\pm$~0.02)$^{\circ}$  is compatible with the previously reported extent of (0.48~$\pm$~0.03)$^{\circ}$. Along the 120$^{\circ}$ direction (the minor axis in Figure~\ref{overviewmap}) the sum of a broad Gaussian component and a narrow Gaussian component describes the profile better. The two-component fit improves
the $\chi^2/dof$ from 48/15 to 5/12. The width of the broad component is (0.60~$\pm$~0.04)$^{\circ}$. The narrow component has a width of (0.12~$\pm$~0.02)$^{\circ}$. The extent of the minor axis profile was underestimated in the previous observations (\citealp{velaHESS}).

 A test for a point source component in VHE $\gamma$-rays from the pulsar position was conducted by the H.E.S.S. collaboration with the first VHE observations of Vela~X (\citealp{velaHESS}). This is repeated here with the new data. The $\gamma$-ray excess found at the pulsar position as a part of the extended emission (Figure~\ref{excessmap80}) has become more significant with respect to the previous observations. However, after subtraction of a Gaussian fit to the surface brightness within 0.1$^{\circ}$ from the pulsar the estimate of the residual $\gamma$-ray flux does not exhibit any significant point-like emission. Nevertheless, it should be noted here that the procedure of fitting the complex Vela~X morphology with a simple Gaussian profile may bias the search.

\section{Discussion}
\label{sec:disc}
The new TeV data presented here allow us to readdress the question of
whether the signal detected with H.E.S.S. corresponds to the X-ray cocoon as was thought in the past, or to the larger radio nebula. For a comparison with the X-ray features of Vela~X a count map from the ${\it ROSAT}$ satellite including events above 1.3 keV, as presented by \cite{aschenbach2}, is used~\footnote{Archival data: http://www.mpe.mpg.de/xray/wave/rosat/index.php}. In the X-ray count map, foreground emission (which may be mainly of thermal origin) covers the entire Vela~X region with a width of $\sim$ 2.5$^{\circ}$ in the east-west direction and spanning a region up to the edge of the SNR shell about 4$^{\circ}$ to the north of the pulsar. An emission level of 400 counts per deg$^2$, morphologically compatible with the larger Vela SNR shell (\citealp{aschenbach2}), is subtracted from the X-ray map. 
Residual thermal emission can still be present after the subtraction of a constant emission level and furthermore the map used for comparison may still  miss fainter emission. As already mentioned the extension of the X-ray nebula is still an open question.

For comparison with data in the radio range, a survey of the Southern Galactic plane with the ${\it Parkes}$ telescope at 2.4~GHz is used (from archival data; \citealp{duncan}). The half-power beam width of the final radio map is 0.17$^{\circ}$, which is slightly broader than the H.E.S.S. PSF. 

In Figure~\ref{3wl} the X-ray cocoon excess map from ${\it ROSAT}$ data, smoothed with a Gaussian function of 0.07$^{\circ}$ rms width, and the ${\it Parkes}$ radio map are shown together with VHE $\gamma$-ray surface brightness contours. Here we have shown that there is a clear VHE $\gamma$-ray signal up to 1.2$^{\circ}$. In contrast to the former interpretation dating back to 2006, the morphology of this VHE $\gamma$-ray emission is intermediate between the X-ray cocoon and radio morphologies. This is also shown in Figure~\ref{slices} where the average brightness profiles of the X-ray map (blue line), excluding the bright compact emission within a circle of radius 6$^{\mathrm{\prime}}$ around the pulsar position, and the radio map (red line) are compared to the H.E.S.S. $\gamma$-ray surface brightness profiles (filled black circles). The profiles through the central position are aligned parallel to the major axis of Vela~X (30$^{\circ}$ anticlockwise from the north), and parallel to the minor axis, respectively. 
Along the minor axis (right-hand panel of Figure~\ref{slices}) the shape of the X-ray profile matches the central part of the VHE component. Hence, the putative narrow component may be a counterpart of the X-ray cocoon. At larger distance from the centre, however, the VHE minor axis profile has a broader extent than the X-ray cocoon. The radio profile is in reasonable agreement with the VHE profile along the two axes of Vela~X. The X-ray cocoon emission peaks close to the pulsar outgoing to the south (at distance $>$ 0.3$^{\circ}$), whereas the VHE and radio profiles do not. The VHE and radio emission maxima coincide (according to Figure~\ref{slices}). This result raises doubts about the simple association of X-ray and VHE emission, and on the contrary suggests an association of the VHE and radio emission. 

Typical approaches in modelling PWNe and in particular the favoured leptonic models discussed for Vela~X assume effective parameters, i.e. the magnetic field and plasma density, averaged within one zone for the entire nebula volume and two distinct populations of electrons (\citealp{zhang}; \citealp{dejager2}). Assuming constant conditions implies that the spatial morphology of the X-ray and VHE $\gamma$-ray emission should match each other, as the emission is generated from the same leptonic plasma in synchrotron and IC processes. 
In order to test the assumption of constant conditions throughout the PWN and that both distinct populations of electrons contribute at the VHE emission, we investigate whether it is possible to describe the morphological structure measured in one wavelength regime with a linear combination of observations in other wavebands. More specifically, we consider a linear combination of the radio nebula $m_r(\bold{x})$ and the X-ray map $m_x(\bold{x})$ as a simple model of the VHE flux $S_m(\bold{x})$:

\begin{equation}
\label{linear}
S\!_{m}(\bold{x}) = a \times m_r(\bold{x}) + b \times m_x(\bold{x}),
\end{equation}
where $a$ and $b$ are location-independent linear coefficients and $\bold{x}$ is the position vector on the sky. 
%Due to the $B^2$ dependence of the synchrotron radiation, the model only works if the effective magnetic field is constant over the whole PWN volume. 
By fitting $S\!_m(\bold{x})$ to the surface brightness map the coefficients $a$ = (1.3~$\pm$~0.1)~$\times$~$10^{\mathrm{-12}}\,\mathrm{cm}^{\mathrm{-2}}\,\mathrm{s}^{\mathrm{-1}}\,\mathrm{deg}^{\mathrm{-2}}$ (Jy/Beam)$^{-1}$ and $b$ = (3.1~$\pm$~0.3)~$\times$~$10^{\mathrm{-15}}\,\mathrm{cm}^{\mathrm{-2}}\,\mathrm{s}^{\mathrm{-1}}$ $\times$ counts$^{\mathrm{-1}}$ are obtained. This result suggests that $\simeq$65\% of the integrated $\gamma$-ray surface brightness is accounted for by a ``radio-like'' component and $\simeq$35\% by an  ``X-ray-like'' component.
In Figure~\ref{slices} the linear combination (black line) of the profiles of the X-ray map (blue line) and the radio map (red line) is presented and compared to the H.E.S.S. $\gamma$-ray profiles (filled black circles). 
The compatibility of the two-component model can be interpreted in two ways: i) the VHE morphology is intermediate between the X-ray and radio morphologies, plausible given the intermediate energies of the electrons probed by VHE IC emission, or ii) simply that both the radio nebula and X-ray cocoon features have VHE counterparts. Furthermore the overall consistency between IC and synchrotron emission suggests a relatively uniform B-field in the nebula on large spatial scales (outside filamentary structures), consistent with previous findings (\citealp{dejager2}; \citealp{fermivela}).

The emission profiles show that along the major axis (i.e. along the main X-ray filament within the cocoon) the VHE $\gamma$-ray emission has a maximum at a radial distance of $\sim$0.7$^{\circ}$ (corresponding to about 4 pc distance) from the X-ray maximum at the pulsar position. The difference in size between the $\gamma$-ray and X-ray cocoon emission regions can be attributed to the difference in energy of the electrons responsible for the radiation. For a few $\mu$G magnetic field only the most energetic electrons (many tens of TeV) emitted from the pulsar produce X-rays in the keV range. The electrons emitting VHE $\gamma$-rays, with characteristic cooling times of some kyr, are accumulated over a few kyr, convected over a larger distance and follow preferentially the $B$-field lines of the cocoon filament (\citealp{dejagerA} and \citealp{hinton}).
 The outer region of the $\gamma$-ray emission seen by H.E.S.S. has a limited X-ray counterpart (mainly along the minor axis) while it is in good agreement with the radio emission. This implies that a second electron population which contributes to the radio emission could be also at the origin of VHE emission. Higher-resolution radio observations at 843~MHz performed with ${\it MOST}$ (\citealp{bock}) have shown a network of arc-like and loop-like filaments distributed over the Vela~X region. They are interpreted as regions of enhanced matter density and/or localized sites of compression (i.e. amplification) of the magnetic field (\citealp{gvara}). The correspondence between enhanced radio synchrotron emission filaments and VHE $\gamma$-rays emission is difficult to establish because of the limited angular resolution of H.E.S.S. compared to ${\it MOST}$.  Furthermore, a correspondence between VHE $\gamma$-ray emission seen by H.E.S.S. and the radio filamentary structures with larger field strength ($>$5$\mu$G) is difficult to explain since the electrons there would have even lower energy. 

Under the assumption that two different electron populations contribute both at VHE one can expect very different VHE spectra for the two components while this situation is not verified. The formerly reported detection of VHE $\gamma$-ray emission in the inner test region has been confirmed in this work, including the spectral shape. The spectral shape of the VHE photon flux from Vela~X follows a hard power law (${\it \Gamma}$=1.32) with an exponential cutoff ($E_\mathrm{cut}$=14 TeV). No significant spectral variations were found within Vela~X and in particular between the cocoon and the outer part of the VHE emission region. Therefore, the first clear measurement of a peak in the spectral energy distribution (of very likely IC origin) at VHE energies applies not only to the cocoon but to the overall larger Vela~X region. The absence of significant spectral variation of the non-thermal emission is surprising in most of the scenarios that have been developed for this object. The lack of an electron cooling signature suggests that particle escape and/or reacceleration may be playing a role. It could have an implication: the involved electron population originating from the accumulation of particles over several thousand years of the pulsar lifetime is modified by energy-dependent diffusive escape and possible stochastic re-acceleration. 
An injection of new electrons accelerated by the pulsar would explain a possible hardening of the spectral index near the pulsar, but this evidence is not established since the measured variation of the spectral index is not statistically significant. For the same reason the extent of Vela~X emission was studied as a function of the energy: no definitive evidence of a radial energy-dependence of the source around the pulsar position is established. 

No significant point-like emission at the pulsar position was found after subtraction of a Gaussian fit to the surface brightness of the extended emission. However, a significant brightness of the order of $10^{\mathrm{-11}}\,\mathrm{cm}^{\mathrm{-2}}\,\mathrm{s}^{\mathrm{-1}}\,\mathrm{deg}^{\mathrm{-2}}$ is present at the pulsar position. A photon index of 1.35 $\pm$ 0.12 was obtained from a power-law fit in the energy range of 0.75 to 10 TeV within 0.2$^{\circ}$ from the pulsar position. Furthermore, the profile along the major axis (Figure~\ref{slices}, left-hand panel) also indicates that the emission fades smoothly, extending to about 0.5$^{\circ}$ north of the pulsar.

\begin{figure}[!t]
  \centering
\includegraphics[width=1\linewidth]{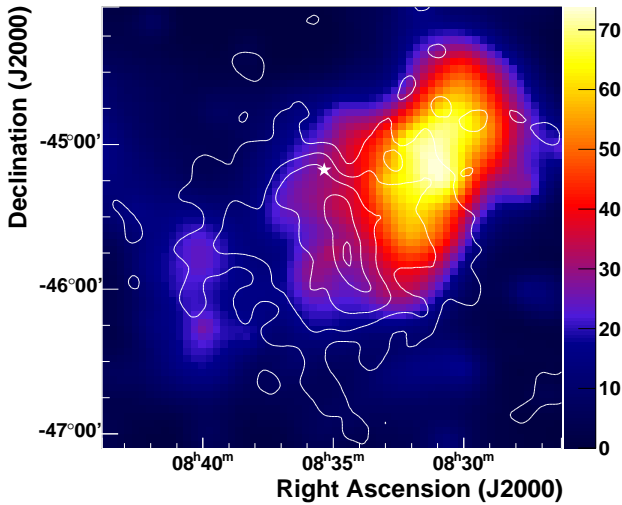}
  \caption{{\it Fermi}-LAT Test Statistic (TS) map (adapted from \citealp{fermivela}) of off-pulse emission in the Vela region above 800 MeV. The measure of the statistical significance for the detection of a $\gamma$-ray source in excess of the background is $S\approx\sqrt{\mathrm{TS}}$. White contours correspond to VHE $\gamma$-ray surface brightness of 0.3, 0.6, 1, 1.6 and 1.9 $\times$ $10^{\mathrm{-11}}\,\mathrm{cm}^{\mathrm{-2}}\,\mathrm{s}^{\mathrm{-1}}\,\mathrm{deg}^{\mathrm{-2}}$.}
  \label{fermiplot}
\end{figure}

The two distinct electron populations (according to the model by \citealp{dejager2}) were also considered by the {\it Fermi}-LAT collaboration to interpret the {\it Fermi}-LAT Vela~X observations. The first component responsible for the IC photon emission in the MeV/GeV energy range and the synchrotron emission at radio wavelengths is considered to interpret the excess seen by the {\it Fermi}-LAT in the energy range from 0.8 to 20 GeV observed in a region of the sky outside the Vela~X cocoon (called $halo$) (\citealp{fermivela}). A second more energetic leptonic component would emit VHE $\gamma$-rays through IC scattering and X-rays of synchrotron origin. Two further pieces of evidence were presented to support this interpretation: 1) the SED is compatible with expectations based on the model and the morphological agreement between the {\it Fermi}-LAT (GeV) and ${\it WMAP}$ (radio) data; 2) negligible VHE $\gamma$-ray emission from the same region of the sky as deduced by the previous H.E.S.S. observations of Vela~X. In this scenario the Vela~X emission detected by H.E.S.S. would originate only from the first leptonic component of the \cite{dejager2} model. The results of the re-observation of Vela~X with H.E.S.S. as described in this work reveal on the contrary that the two-population model needs to be revised since the region explored by {\it Fermi}-LAT (Figure~\ref{fermiplot}), called {\it halo} by the authors, is characterised also by a significant VHE $\gamma$-ray emission. 

One alternative hypothesis to interpret the GeV and the VHE spectra is to infer different magnetic fields present in the region. In line with this hypothesis it is interesting that the time-dependent model recently proposed by \cite{hinton} uses the present-day B-field in the extended radio nebula and a second B-field for the cocoon as free parameters. Unfortunately, the possibility to quantitatively support this hypothesis by means of a multiwavelength spectral energy distribution comparing the GeV and VHE $\gamma$-ray emissions is limited by the fact that the morphologies of the explored regions vary between the different wavebands as shown by the results of the work here reported. Finally, in this context it is worth recalling that the scenario is even more complicated when considering data from both ${\it AGILE}$ and {\it Fermi}-LAT. When combined with the H.E.S.S. data, both show multiple peaks of potential IC origin in the SED, but the results obtained with the two instruments are incompatible with regard to morphology and spectra. The photon index measured with ${\it AGILE}$ ($\sim$1.7) is harder than the one found by {\it Fermi}-LAT ($\sim$ 2.4). Furthermore, in contrast to the emission detected by {\it Fermi}-LAT, the source AGL J0834-4539 detected by ${\it AGILE}$ in the energy range 100 MeV to 3 GeV is positionally coincident with HESS J0835-455 and has a brightness profile similar to the VHE Vela X morphology measured by H.E.S.S. (\citealp{pelli}).

In conclusion, revisiting the Vela~X region with a refined analysis of its VHE $\gamma$-ray emission based on an enlarged data set demonstrates the puzzles often encountered in the interpretation of multiwavelength data when high-energy measurements are involved. At the same time, our study emphasises the relevance of the contributions of present and future VHE $\gamma$-ray IACT systems, with their large field of view and high sensitivity, to the understanding of the physics of PWNe.

\begin{acknowledgements}
The support of the Namibian authorities and of the University of Namibia
in facilitating the construction and operation of H.E.S.S. is gratefully
acknowledged, as is the support by the German Ministry for Education and
Research (BMBF), the Max Planck Society, the French Ministry for Research,
the CNRS-IN2P3 and the Astroparticle Interdisciplinary Programme of the
CNRS, the U.K. Science and Technology Facilities Council (STFC),
the IPNP of the Charles University, the Czech Science Foundation, the Polish 
Ministry of Science and  Higher Education, the South African Department of
Science and Technology and National Research Foundation, and by the
University of Namibia. We appreciate the excellent work of the technical
support staff in Berlin, Durham, Hamburg, Heidelberg, Palaiseau, Paris,
Saclay, and in Namibia in the construction and operation of the
equipment. 

\end{acknowledgements}

\end{document}